\documentclass{article}
\usepackage[margin=2.9cm]{geometry}
\usepackage{mathptmx}
\usepackage{amsmath,amssymb,latexsym}
\usepackage{amsfonts}
\usepackage{graphicx}
\usepackage{authblk}
\usepackage{algorithm2e}   % package to write an algorithm
\providecommand{\keywords}[1]
{
  \small	
  \textbf{\textit{Keywords---}} #1
}

\newcommand{\bx}{\mathbf{x}}    

\begin{document}

\title{Spatio-temporal regularization of global ocean waves obtained from satellite and their graphical representation}

\title{\bf Spatio-temporal regularization of global ocean waves obtained from satellite and their graphical representation}
\author[1]{Heitor Perotto}
\author[2]{Leandro Farina}
\author[3]{Nelson Violante-Carvalho}
\affil[1]{\small Institute of Geosciences, Federal University of Rio Grande do Sul, Porto Alegre, RS, Brazil}
\affil[2]{Institute of Mathematics and Statistics, Federal University of Rio Grande do Sul, Porto Alegre, RS, Brazil}
\affil[3]{Program of Ocean Engineering, Federal University of Rio de Janeiro, Rio de Janeiro, RJ, Brazil}

\maketitle

\begin{abstract}
An algorithm for representing irregular satellite data obtained from advanced synthetic aperture radar (ASAR) uniformly in space and time is introduced and comparison with WAVEWATCH III model (WW3) data is carried out. Swell global data for 2007 are analysed using statistical parameters such as bias and scattering index. Satellite data showed a great spatial uniformity compared with WW3 model.
Results show underestimation of swell height by the model in the tropics and at specific higher latitude regions.
\end{abstract}


\section{Introduction} \label{intro}
In the fields of physical oceanography and ocean engineering, the advent of satellite remote sensing has substantially increased our understanding of several synoptic and sub-synoptic phenomena.
The spatio-temporal sparseness and irregularity of satellite data 
could be regarded as the most severe limitation, as the satellite takes typically several days to return to the same point. On the other hand, surface ocean wave measurements by satellite constitute a remarkable source of data for describing wave climate, improving forecast and ocean and coastal monitoring. 
Thus, representing actual wave variables such, as height, in a regular grid is a problem of great interest. Assigning these data, for a certain time interval, on a regular grid will typical produce a map with {\em missing data}. Adding values, interpolating and taking averages are simple alternatives to solve the problem. More sophisticated methods, such as empirical orthogonal functions~\cite{alvera2005,alvera2009}, singular spectrum analysis~\cite{ghil2006} and neural networks~\cite{barth2020} have been proposed for geophysical data sets, although not specifically for ocean wave variables.

Investigations about wave climate have increased and satellite data have been increasingly employed. Satellite observations have a global coverage and also provide information with a high level of accuracy. Along with satellite data, numerical modelling can be used as a tool for providing numerical predictions~\cite{Reguero2012,laugel2014}. Despite the irregular spatial and temporal coverage of satellite wave data, these records, along with in situ ones, are supposedly the closest measurements to the true sea state. The results of a model for forecasting or hindcasting wave fields provide a regular, homogeneous field in time and space. However, they are subject to many uncertainties, such as the quality of wind input and the model physics and configuration. Thus, a comparison of satellite data against model data model becomes very useful to determine the real sea state.

Young~\cite{Young1999}  examined a data set covering a 10-year period acquired from a combination of satellite data and model predictions obtained in order to establish  wind and wave global climates. The results were presented in terms of average monthly statistics for significant wave height, peak and average wave period, wave direction and wind speed and direction. The results clearly showed zonal variation for the wind speed and wave height, to extreme conditions that occur at high latitudes. It was observed that the swell generated from storms in the Southern Ocean penetrates across the Indian Ocean, South Pacific and South Atlantic. During the southern hemisphere winter, the swell still enters the North Pacific. It was also found that the western side of most continents have noticeably more intense wave climates than the east side; this is a result of the generation of longer fetches that exist on the West Coast and the direction of propagation of prevailing winds. Later, Young et al.~\cite{Young2011} extended the work using a 23-year database of calibrated and validated satellite altimeter data to establish global trends in wind speed and wave height.

A feasibility study to use wind speed and significant wave height measurements of simultaneous sources of scatterometer and altimeter satellite to observe the pattern of spatial and seasonal distribution of the dominant areas of swell and wind in all oceans has been proposed by~\cite{Chen2002}. Two normalised indices related to energy were presented. Based on these, global probability and intensity statistics for swell and wind were obtained. Regions with intense wave growth were observed in the Pacific, Northwest Atlantic, Southern Ocean and the Mediterranean Sea. The observed seasonality is somewhat different between the climates of swell and wind.

An analysis using a set of satellite altimeter data was carried out by Izaguirre et al.~\cite{Izaguirre2011}, which provided a global comprehensive level for the year of 1992. An analysis of non-stationary extreme values, modelling the seasonal and inter annual variations was employed to characterise extreme significant wave height.

Ren et al.~\cite{Ren2011} compared six years of ENVISAT satellite ASAR spectrum data with one-dimensional wave spectra obtained from 44 buoys in the Northern Pacific. They found that the ASAR wave spectra tend to notably underestimate the significant wave heights at high wind speeds and overestimate them at low wind speeds.

Reguero et al.~\cite{Reguero2012} used a global set of wave data from 1948 to 2008. Employing a calibration method, the data set have been fixed based on altimeter data from 1992 to 2008. The quality of the corrected results  were compared with measurements from buoy and satellite altimeter. The diagnostic statistics showed an agreement both with scatter data as the statistical distribution of wave heights, indicating that the reanalysis adequately reflects the characteristics of the waves from satellite data between 1992 and 2008.

ENVISAT level 2 (ASA \_WVW\_2P) wave spectra data were compared with other altimeter and buoys measurements, located in the Gulf of Mexico and in the Pacific Ocean, and with wave models by Li and Saulter~\cite{Li2012}. This indirect comparison was made for 14 months, from November 2007 to December 2008. The ASAR swell data proved to be consistent with the model and buoy data.

Stopa et al.~\cite{Stopa2016} used synthetic aperture radar data to identify swell sources and trajectories, allowing a statistically significant estimation of swell dissipation. They employed the whole ENVISAT mission (2003 - 2012) to find suitable storms with swells (with periods $T$ such that $13 < T < 18$ s) 
that were observed several times along their propagation. The analysis reveals that swell dissipation weakly correlates with the wave steepness, wind speed, orbital wave velocity, and the relative direction of wind and waves. Although several negative dissipation rates are found, there are uncertainties in the SAR-derived swell heights and dissipation rates. An acceptable range of the swell-dissipation rate is $-0.1$ to $6 \times 10^{-7}$ $\mbox{m}^{-1}$ with a median of $1\times 10^{-7}$ $\mbox{m}^{-1}$.

One of the shortcomings of data obtained by satellite is the irregular spatial and temporal distribution inherent to these. In the above works cited (see also 
~\cite{woolf2002,hemer2010}) averages are used to overcome this limitation and to allow climate investigations and comparisons with other data.
In this work, we present an algorithm that reallocate the wave data in a uniform grid {\em without the use of averages or means}.  Thus, only actual, pontwise measured values are plotted and analysed. This is the main difference of our approach with respect to previous works on this topic. The algorithm is composed of two main procedures; the first assigns measurements of sattelite swell heights (not mean values) to the computational cells of a wave forecast model. This is done for data in a certain period of time such that most of the model`s grid cells are filled by the sattelite data. We found empirically that the optimum period of time is one month. If two sattelites measurements for the same cell are found, the largest value is selected. The second main procedure is responsible for making both sets of data (model`s and sattelite) uniform in the time domain. This is done by selecting the model`s data which are simultaneously in the same cell as the previously assgined sattelite data and closest in time, among the values in the model`s discretized time domain. As a result, the algorithm produces actual values of both sattelite and model's data represented in the same regular submesh of the original model's domain.

Another important aspect of the approach is related to how the data are plotted: the sattelite data are plotted exactely at their original coordinates while the model`s data are placed at the center of the cell they belong to. These values are graphically a colored point/pixel at the two-dimensional map.

A Synthetic Aperture Radar (SAR) inboard satellites can be used to measure directional wave spectrum. In its so called wave mode, small areas of around 5 by 5 km are imaged along the satellite ground track, yielding thousands of wave spectra per orbit with global coverage. Our knowledge about wave climate and wave propagation has increased substantially since the launch of ERS-1 in the early 1990s (see a thoroughly discussion about the SAR wave mode in \cite{escidoc:2039578}).
In this work we analyzed swell heights in a global domain for the whole year of 2007. The data used was the directional wave spectra (ASA \_WVW\_2P) obtained by the ENVISAT satellite of the European Space Agency (ESA) through the ASAR sensor. These data were compared with WAVEWATCH III (WW3) model data, developed  for the same period. 

An important point to stress is that validations of wave models against ASAR wave spectra are very rare to appear in the literature. An important issue we aim to help tackle is that the very few published articles try to compare the wave model output with ASAR wave spectra with coincident measurements along the satellite track, within a few tens of kilometers and a few tens of minutes. To the best of our knowledge this is the first attempt to validate an ensemble of ASAR wave measurements against spatially and temporally gridded model data. The challenge to do so is related to the complexity to put together such unevenly time/spaced data, which is proposed in our paper. With this kind of approach a more elaborated spatial picture of the comparison ASAR/wave model data emerge.

In the next section we describe the data set used. In section~\ref{sec:algo}, we present details of the algorithm for the uniform data representation while 
the overall results on the wave fields are presented and discussed in section~\ref{sec:results}. Concluding remarks are given in section~\ref{sec:end}.

% *********************************************************
\section{Satellite and model wave data} \label{sec:data}
% *********************************************************
% -------------------------
\subsection{Satellite data}
% -------------------------
ENVISAT was launched in March 2002 by the European Space Agency to follow on its predecessors ERS-1 \& 2. Its main objective was to provide atmosphere, ocean, land and ice data to monitor global warming, the degree of air pollution and natural disaster risks. ENVISAT carried ten instruments on board, with optical sensors and active sensors, and among them its main asset, the Advanced Synthetic Aperture Radar (ASAR) sensor. The ENVISAT mission ended on 8 April 2012, following the unexpected loss of contact with the satellite.

The ASAR sensor is an enhanced version of the SAR instrument that was already in operation on board satellites ERS-1 and ERS-2. This sensor has a 10 m long antenna, and operates in five different modes: image mode, mode alternating polarisation wide band mode, monitor mode, and global wave mode. %These modes have resolutions ranging from 25 km to 1 km, specially developed to observe continental areas, ocean areas and polar ice caps. 
The sensor in wave mode produces small  5 km $\times$ 5-10 km images, called {\em imagettes}, of the  sea surface. These images are processed to derive the wave spectrum of the sea surface and hence the wavelength, period, height and direction of the waves. The processing of Level 2 data uses an algorithm developed and implemented by IFREMER/CERSAT~(\cite{ifremer}).

Although the ENIVISAT mission encompasses the years from 2002 to 2012, there are other satellite missions currently operating, such as Sentinel, which provide ocean wave data. We remark that the algorithm to be described is independent of the type of sensor and could be in principle applied to present and future satellite missions. 
% ---------------------------
\subsection{The wave model}
% ---------------------------
The WW3 wave model \cite{Tolman1999} developed by NOAA / NCEP solves the spectral action energy conservation equation. The physical model includes wave growth, dissipation due to wave breaking in the background, refraction, advection and nonlinear interactions.

In our implementation, the WW3 model is used with spatial resolution of 
$0.6\,^{\circ} \times  0.6^{\circ}$ for latitude and longitude respectively and temporal resolution of 3 hours. The simulations are performed in a global domain $ 70\,^{\circ} $ S to $ 70\,^{\circ}$ N and $0$ to $360\,^{\circ}$ E. As input the surface wind components ($U$ and $V$) predictions and ice cover calculated by the NCEP Climate Forecast System Reanalysis (CFSR) are used.

We used data from the spectral partitions provided by the model. The WW3 model has both point and field output options available to provide quantitative descriptions of these individual spectral partitions. The analysis of the wave system described in \cite{Hanson2001,Hanson2004} is applied.

The spectral partitioning method subdivides the directional spectrum into a number of overlapping wave components (partitions), each of which can be characterised by average parameters such as significant height, mean frequency, direction of propagation, and directional scattering.

Regardless of the method used, all the partitioning schemes follow the following logic: isolation of the peaks of energy, identification and combination of peaks of energy, removal of partitions with low energy and statistics of the main parameters of waves.

Hanson and Jensen~\cite{Hanson2004} and Hanson et al.~\cite{Hanson2009} used a MATLAB code to apply the Vincent and Soille~\cite{Vincent1991} algorithm. This code has been transformed to an efficient FORTRAN routine for use in WAVEWATCH III since version 3.11~(\cite{Tolman2009}).
%
% *********************************************************
\section{Uniform spatio-temporal representation} \label{sec:algo}
% *********************************************************
%
The allocation of the satellite data, which is irregular in space, into a regular mesh is carried out by an algorithm with two procedures. The first rearrange the satellite data onto the model regular grid, placing the satellite data closest to its original irregular coordinates. When two satellite measurements are found, within the specified period of time (taken here as one month), in the same cell of the model mesh, the largest measured value is selected. Thus, this procedure sweeps all the satellite data in the globe for the period of one month and assigns every single measurement found in the nearest model's cell. In the present study, this procedure was repeated for every month of the year 2007. The algorithm can be interpreted in simple and didactical terms as an interpolation from the satellite data to a regular grid by piecewise constant to the spatical points of the model grid. The time interpolation is performed likewise but now to the satellite temporal domain.

Let $V_{sat-orig}(\bx,t)$ denote the variable $V$ measured by the satellite at
a point $(\bx,t) = (x,y,t)$ on latitude $x$, longitude $y$ and time $t$. 
Let also $lat(i), lon(j)$ be the latitudes and longitudes of the model's horizontal physical mesh of size $N \times M$; these are the values of the latitude and longitude used in the plots of figures~\ref{fig:fevapr-swell}, \ref{fig:julout-swell}, \ref{oldfig2} and~\ref{oldfig3}.

Thus, the algorithm generates a regular submesh of size $N_b \times M_b$, subset of the original mesh, which can coincide with the model original mesh itself, particularly when the amount of satellite data over one month is
sufficiently large and homogeneously distributed. The first producedure, used for every month studied, is listed with more details in the scheme Procedure~\ref{algo:1} below. 

\renewcommand*{\algorithmcfname}{Procedure}  % Modify the name "Algorithm to Procedure" in the caption

\begin{algorithm*}
 \KwData{Satellite data over a month and model mesh}
 \KwResult{Satellite data $V_{sat-reg}$ arranged on regular submesh}
 \For{ $(\bx,t) \in$ Sat. global domain $\times$ one month}{
  $count(i,j) = false$\; 
  \For{ $i=1,\ldots N-1, j=1,\ldots, M-1$}{
  \If{$x > lat(i)$ and $x \leq lat(i+1)$       
   \mbox{\bf and} $y > lon(j)$ and $y \leq lon(j+1)$}{ 
   \eIf{$count(i,j) = true$}{        
   \eIf(\tcp*[f]{Assign the satellite data to the closest cells}){$V_{sat-reg}(i,j,date(i,j)) < V_{sat-orig}(\bx,t)$}{ 
   $date(i,j) = t$ \tcp*[f]{of the model grid while keeping the satellite time}  
   $V_{sat-reg}(i,j,date(i,j)) = V_{sat-orig}(\bx,t)$}{increment $i ; j$}
}{ $date(i,j) = t$ \;
   $V_{sat-reg}(i,j,date(i,j)) = V_{sat-orig}(\bx,t)$ \;
   $count(i,j) = true$}
}
}
}
 \caption{Reallocate satellite data in regular mesh, where the variables notation above are employed.} \label{algo:1}
\end{algorithm*}

The second procedure compares the dates of the satellite measurements with model dates and selects only the model dates that have a corresponding satellite dates.
% Let $k=1,\ldots,T$ be the dates computed by the model during the month under study.
This is depicted in the scheme Procedure~\ref{algo:2}.

\begin{algorithm*}
 \KwData{Dates where satellite data exists, spatial submesh and model temporal mesh}
 \KwResult{Model data $V_{mod-reg}$ on the submesh and on selected dates}
 \For{ $k = 1,\ldots,T$}{
  \For{ $i=1,\ldots N_b-1, j=1,\ldots, M_b-1$}{
  \If(\tcp*[f]{Assigns the model data at the satellite time}){$ date(i,j) > k$ and $date(i,j) \leq k+1$}{
  $V_{mod-reg}(i,j,date(i,j)) = V_{mod-orig}(i,j,k)$
}
}
}
 \caption{Selects the valid dates on the model data.} \label{algo:2}
\end{algorithm*}

As a result of the algorithm, the variables $V_{mod-reg}$ and $V_{sat-reg}$ can then be compared on the same space-time grid. In our comparison, we have computed swell height.

As mentioned in section~\ref{intro}, the sattelite data are plotted exactely at their original coordinates while the model`s data are placed at the center of the cell they belong to. These values are graphically a colored point/pixel at the two-dimensional map. Thus, they correspond only to time instances with both satellite measurements and model data, coincident in time. There are no monthly means and averages shown.
%
% **************************************************************
\section{Swell Height} \label{sec:results}
% **************************************************************
%
To analyse the data obtained from the ENVISAT ASAR sensor, we carried out a  comparison with data obtained by the WW3 model. 
The swell height was obtained and simulated for the whole year of 2007. 
For the graphical presentation of results, we selected a single month for each of the four seasons, given there is not a significant variation among other months within a season.

Besides the values of swell heights, we also analysed the bias and the scattering index from our dataset. Specifically, we have computed
\begin{displaymath} \label{vies}
V=M-S,
\end{displaymath} 
\begin{displaymath} \label{si}
SI=\frac{\sqrt{\frac{1}{n}\sum_{i=1}^{n}\left(M_i - S_i \right)^2}}{\bar{S}}
\end{displaymath} 
where $V$ is the bias, $SI$ is the scatter index and $n$ is the number of observations. $M$ stands for model and $S$ for satellite
satellite measurements. The bar over $S$ represents spatial average.

Each value of the swell height, as well as the bias and the scatter index is represented by a pixel, with its corresponding latitude and longitude, in the maps of all figures below. 

The satellite wavelength measurements range is 100-400 m, which is equivalent to periods of 8-16 s. Thus we selected waves with corresponding wave periods computed by the model.

\begin{figure*}[h]
    \begin{center}$
    \begin{array}{cc}
\includegraphics[width=6.9cm]{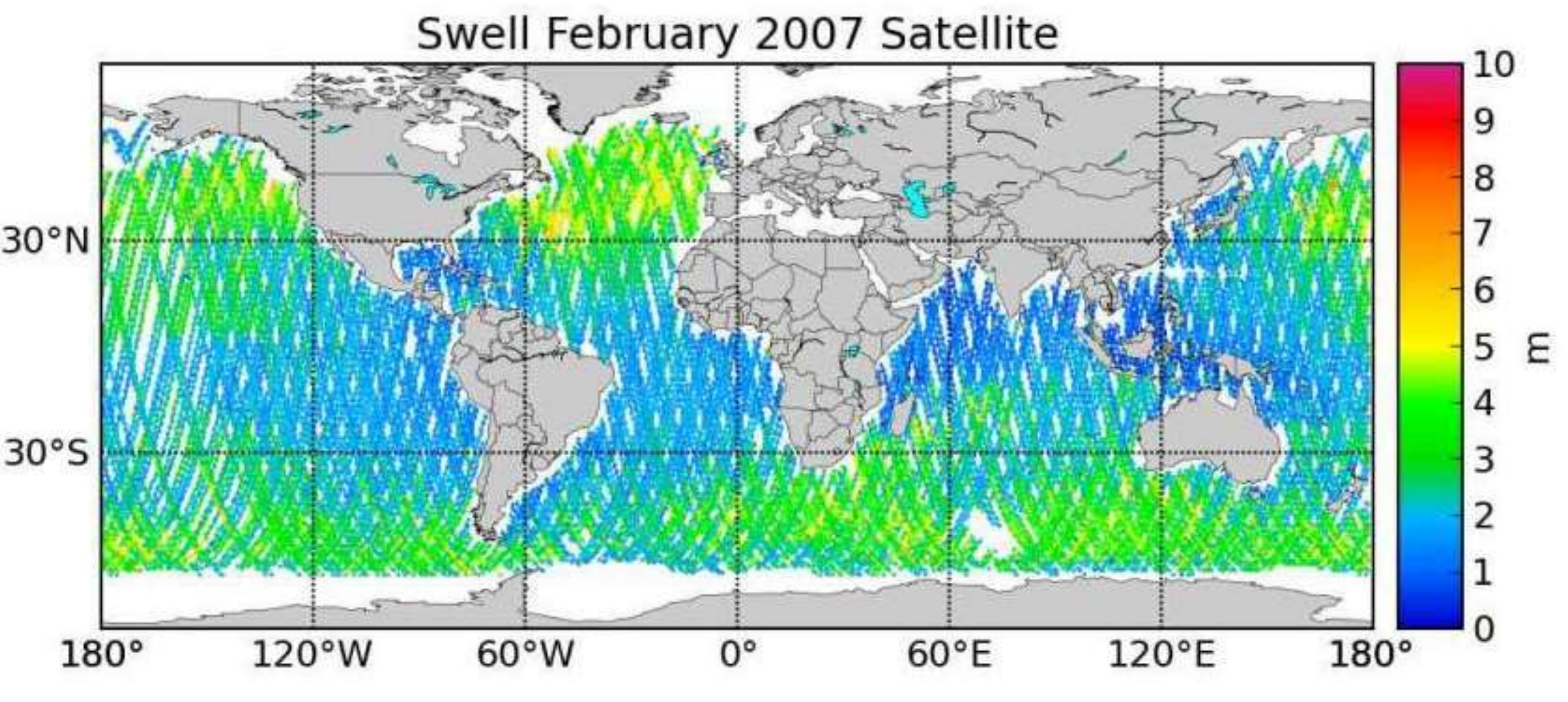} &
    \includegraphics[width=7.0cm]{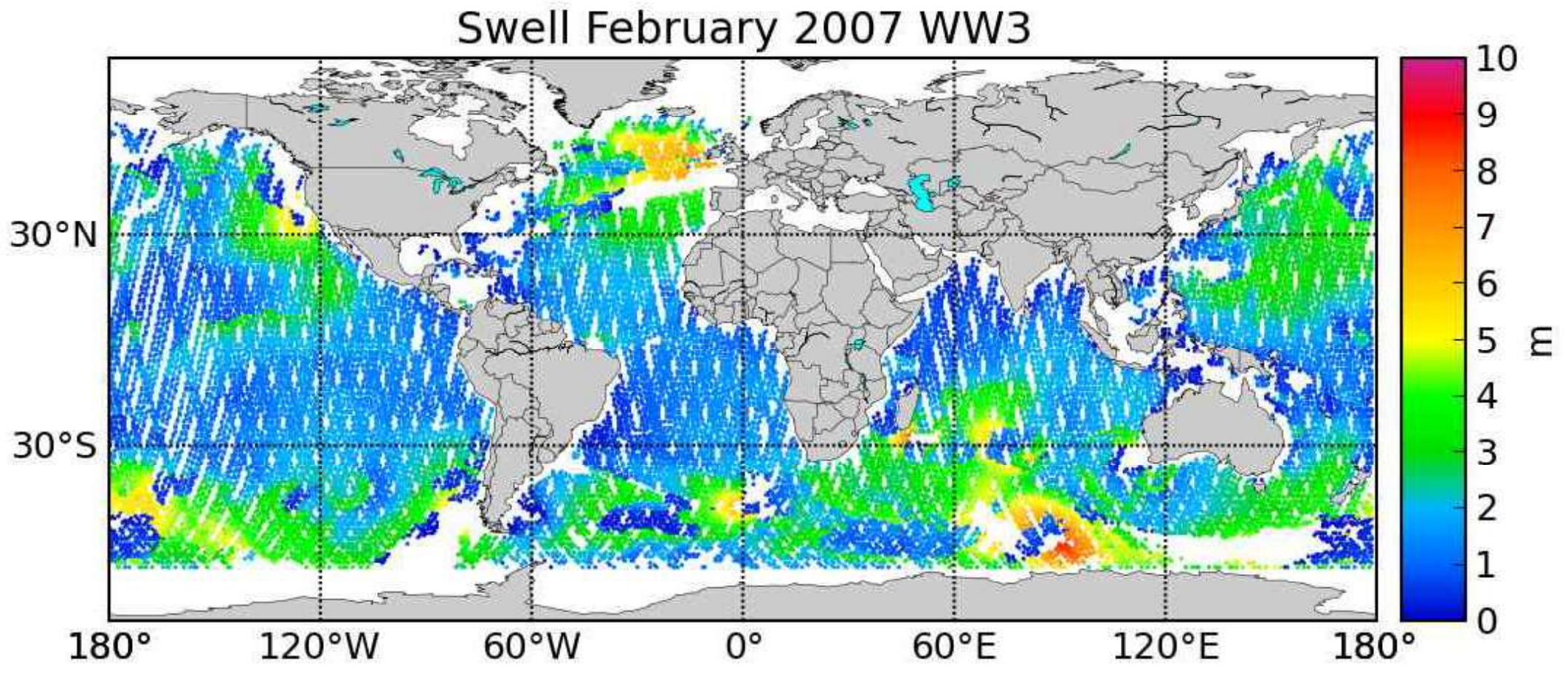} \\
    \includegraphics[width=6.9cm]{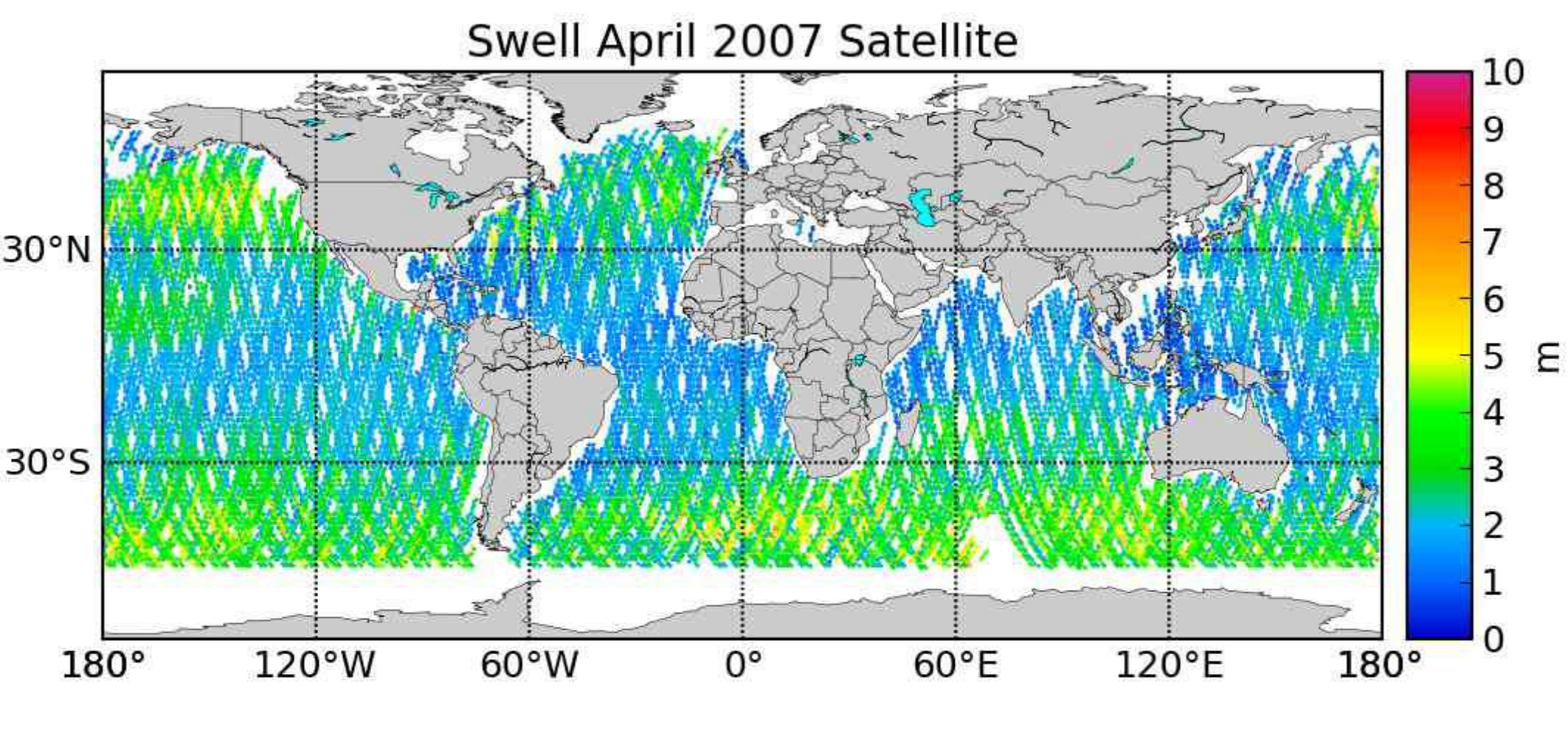} &
    \includegraphics[width=7.cm]{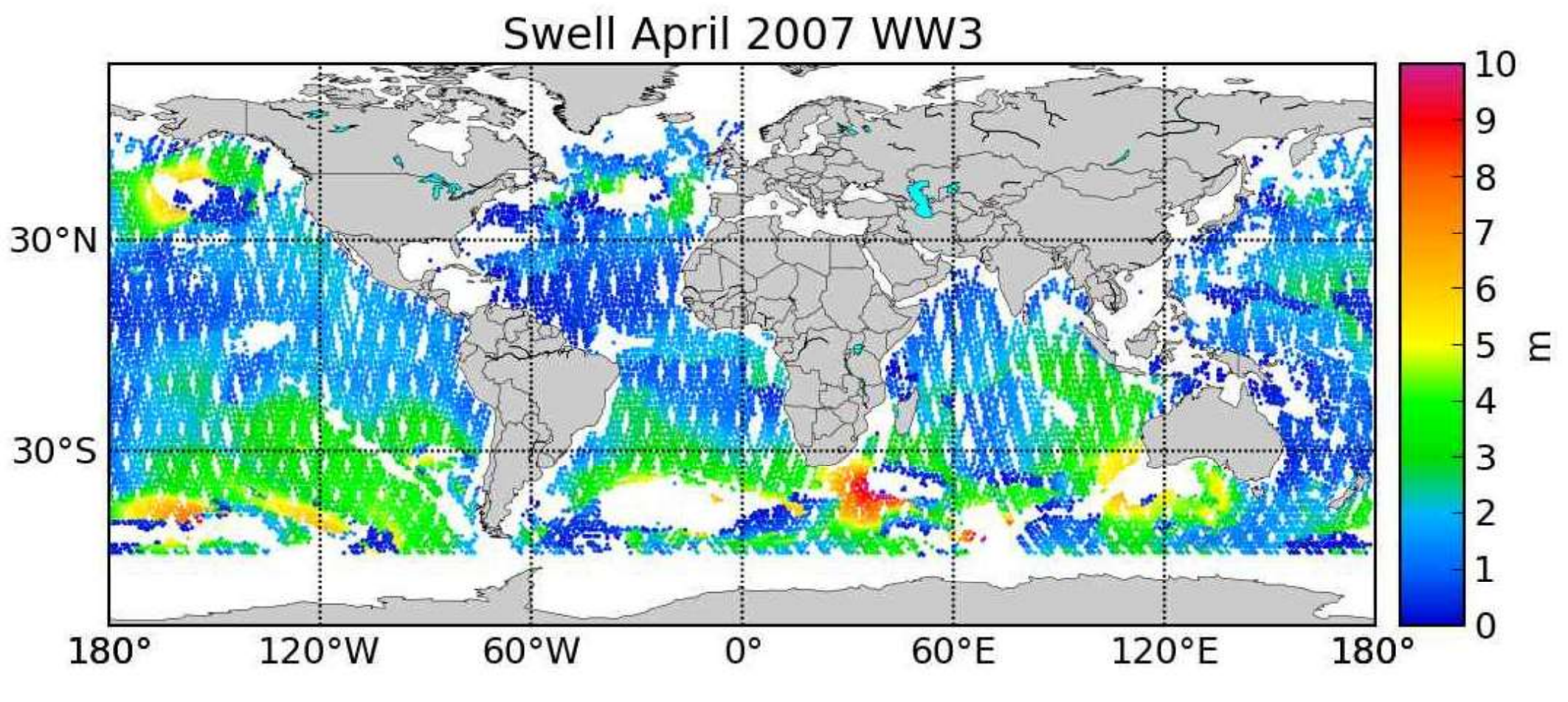}
    \end{array}$
    \caption{Global distributions of swell height in meters by the satellite and model for the months of February and April. 2007.}
    \label{fig:fevapr-swell}
    \end{center}
\end{figure*}

\begin{figure*}[h]
    \begin{center}$
    \begin{array}{cc}
    \includegraphics[width=6.9cm]{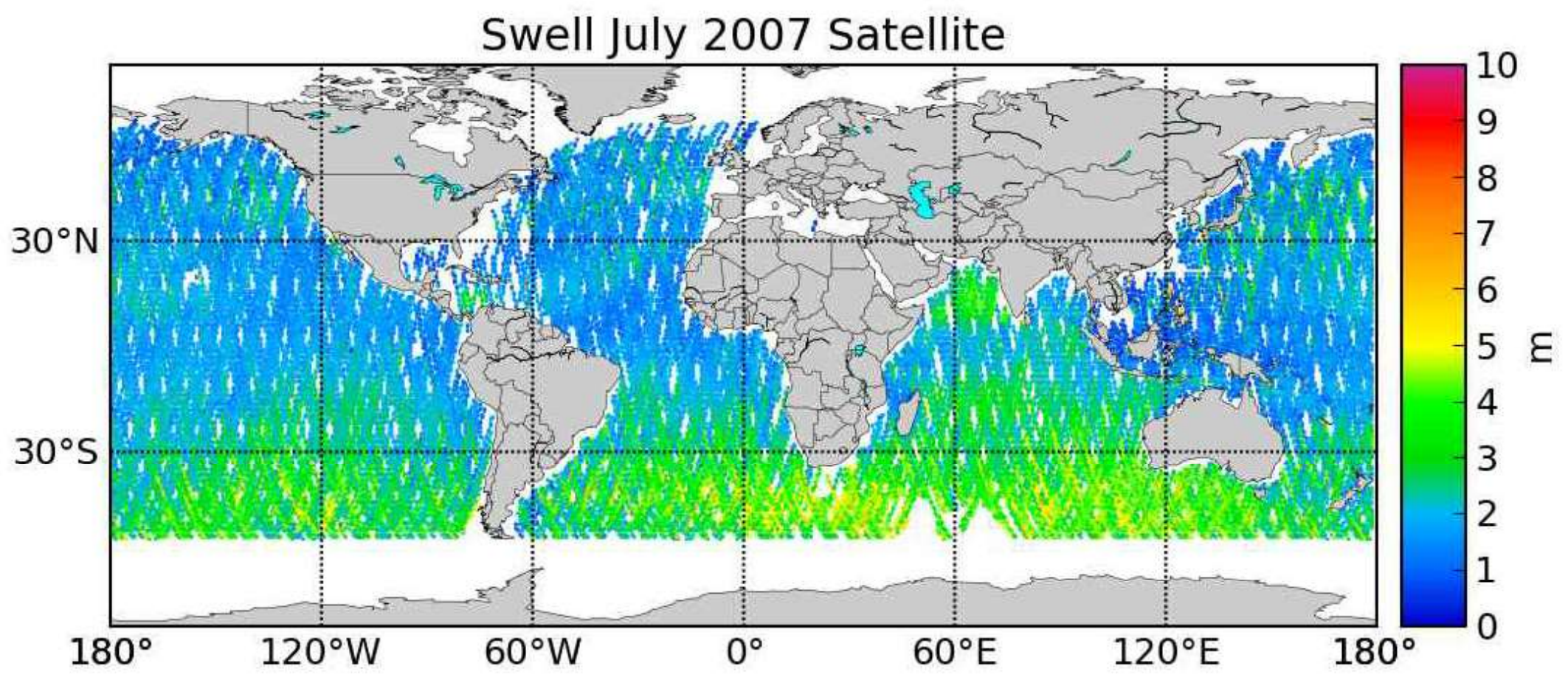} &
    \includegraphics[width=7.0cm]{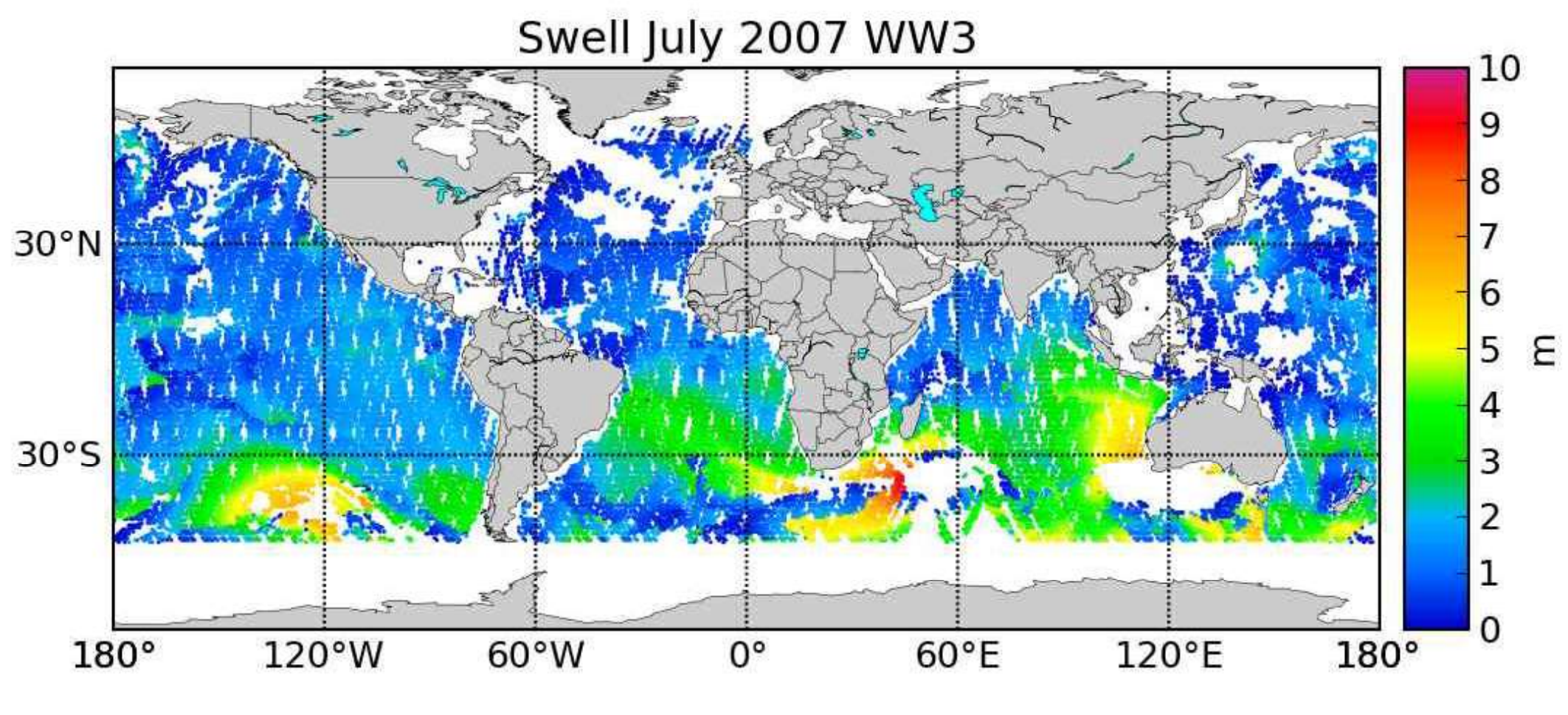} \\
    \includegraphics[width=6.9cm]{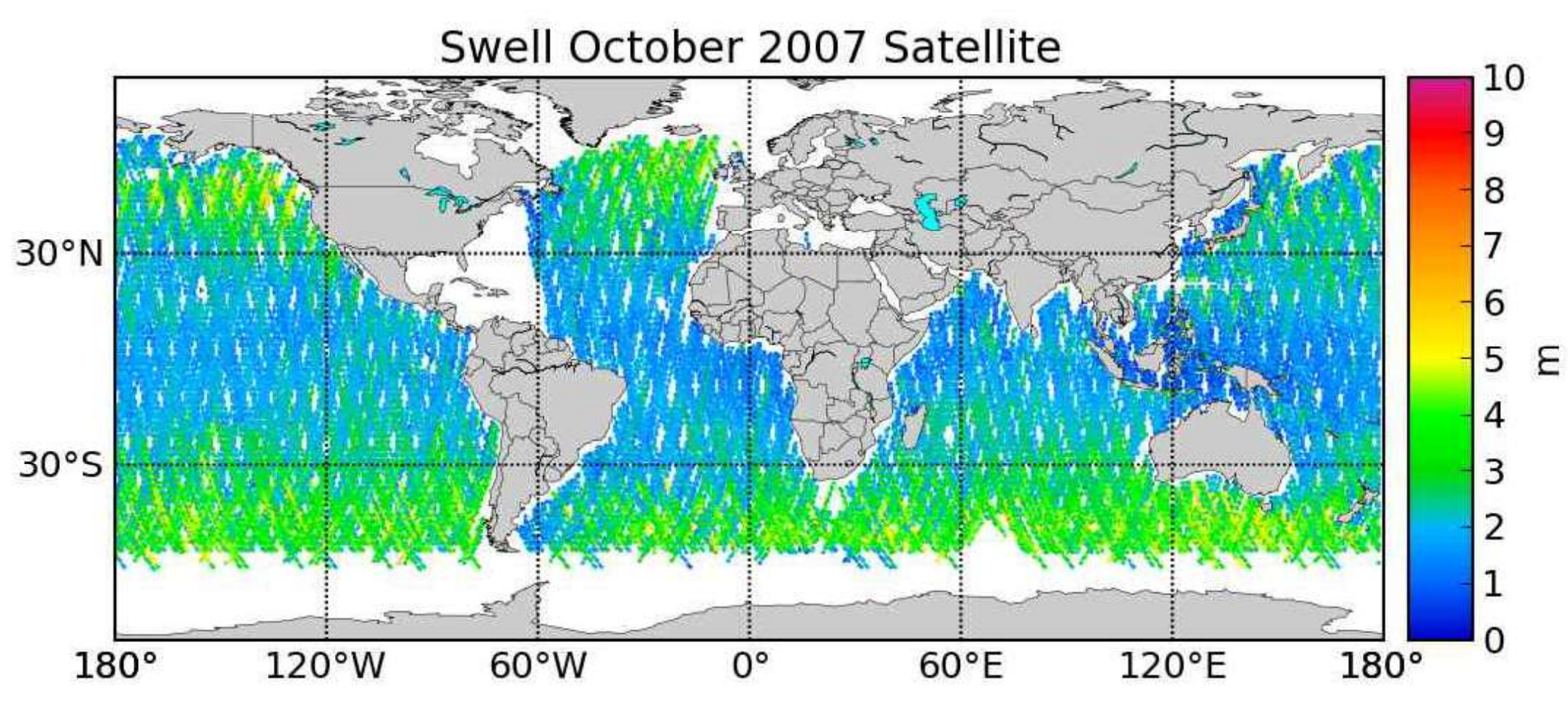} &
    \includegraphics[width=7.cm]{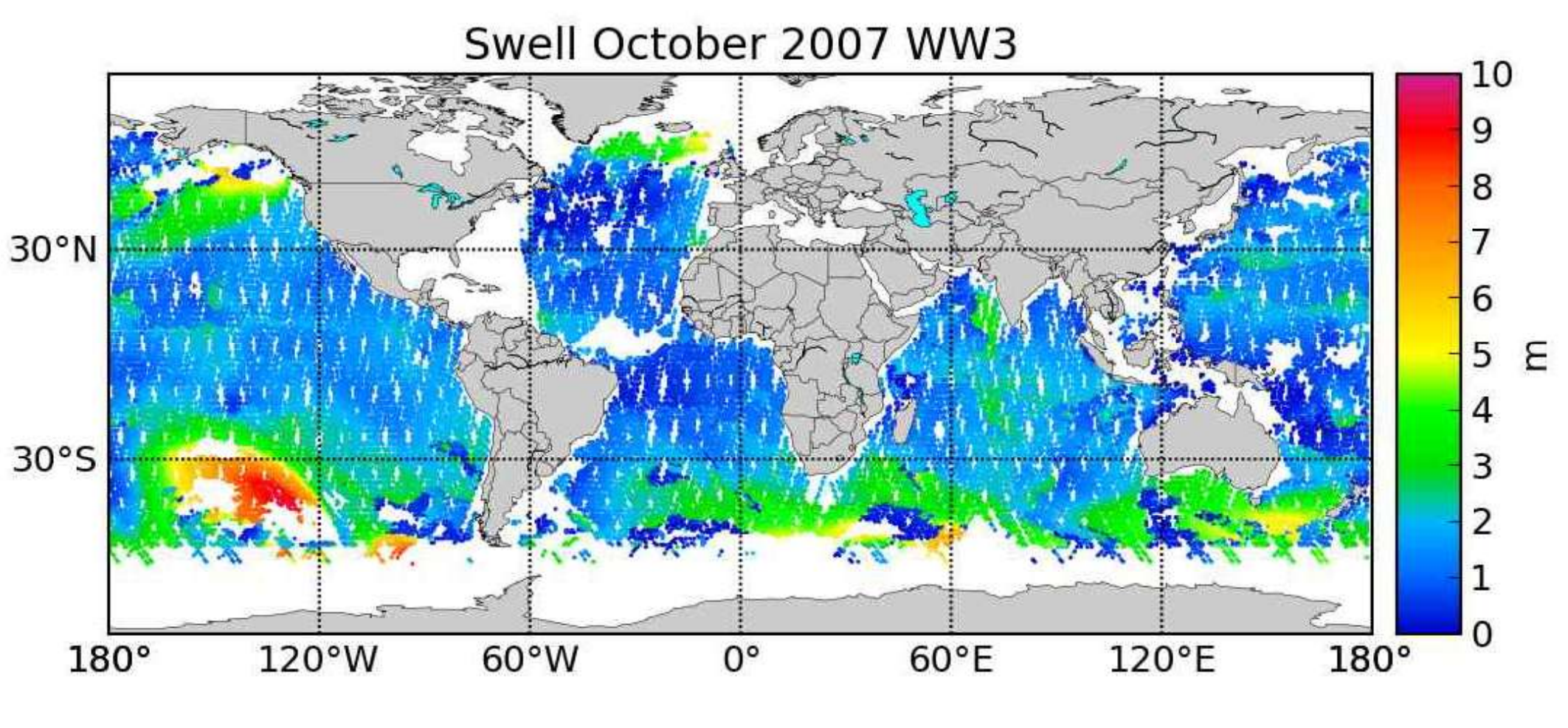}
    \end{array}$
    \caption{Global distributions of swell height in meters by the satellite and model for the months of July and October, 2007.}
    \label{fig:julout-swell}
    \end{center}
\end{figure*}

In all figures in this section a representative month of each season is picked up. As there is not a significant variation among the other months of each season, this choice is justified. It is observed from the satellite and model data in figures~\ref{fig:fevapr-swell} and~\ref{fig:julout-swell} that the largest values of swell height are located near latitude $60\,^{\circ}$ for both the Southern Hemisphere (SH) and the Northern Hemisphere (NH), corroborating with previous findings on ocean wave climate~\cite{Young1999,Izaguirre2011}.
In the months corresponding to winter and fall, in the respective hemispheres, these values intensify.
Analysing both dataset we notice that the SH show a significantly rougher sea state than the NH. In particular, stronger swell fields are observed in the South Pacific, Southern Africa and Southern and Eastern Australia in the autumn and winter of the SH. Important high swell also occur in the Southwestern Atlantic and is seen from the model data of April~(figure~\ref{fig:fevapr-swell}) and in the model and satellite data of July~(figure~\ref{fig:julout-swell}).
These fields are associated with the frequent occurrence of extra-tropical cyclones in the SH, in the fall of the SH. In the NH, the highest waves are found in the North Pacific and North Atlantic during the winter, as shown in figure~\ref{fig:fevapr-swell}.

\begin{figure*}[h]
    \begin{center}$
    \begin{array}{cc}
    \includegraphics[width=6.9cm]{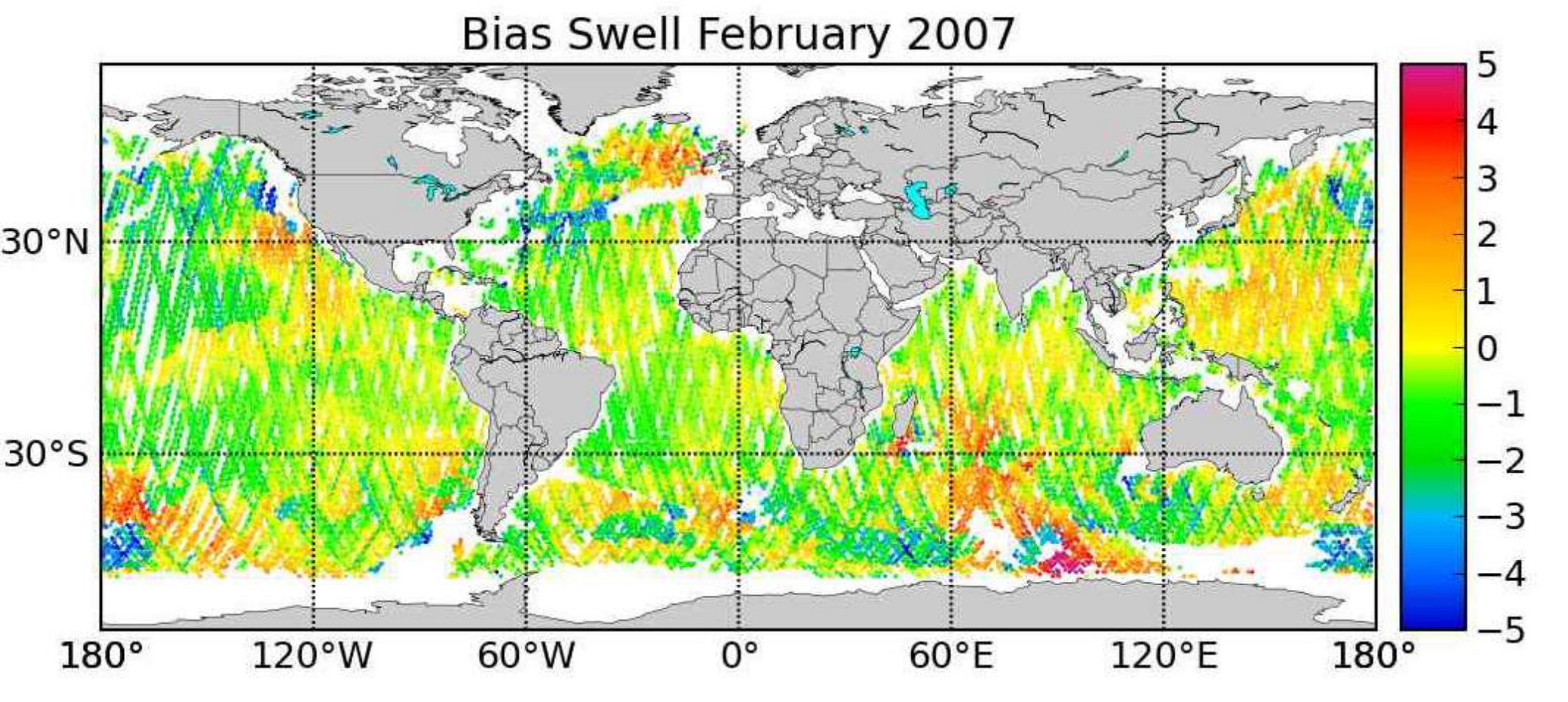} &
    \includegraphics[width=7.cm]{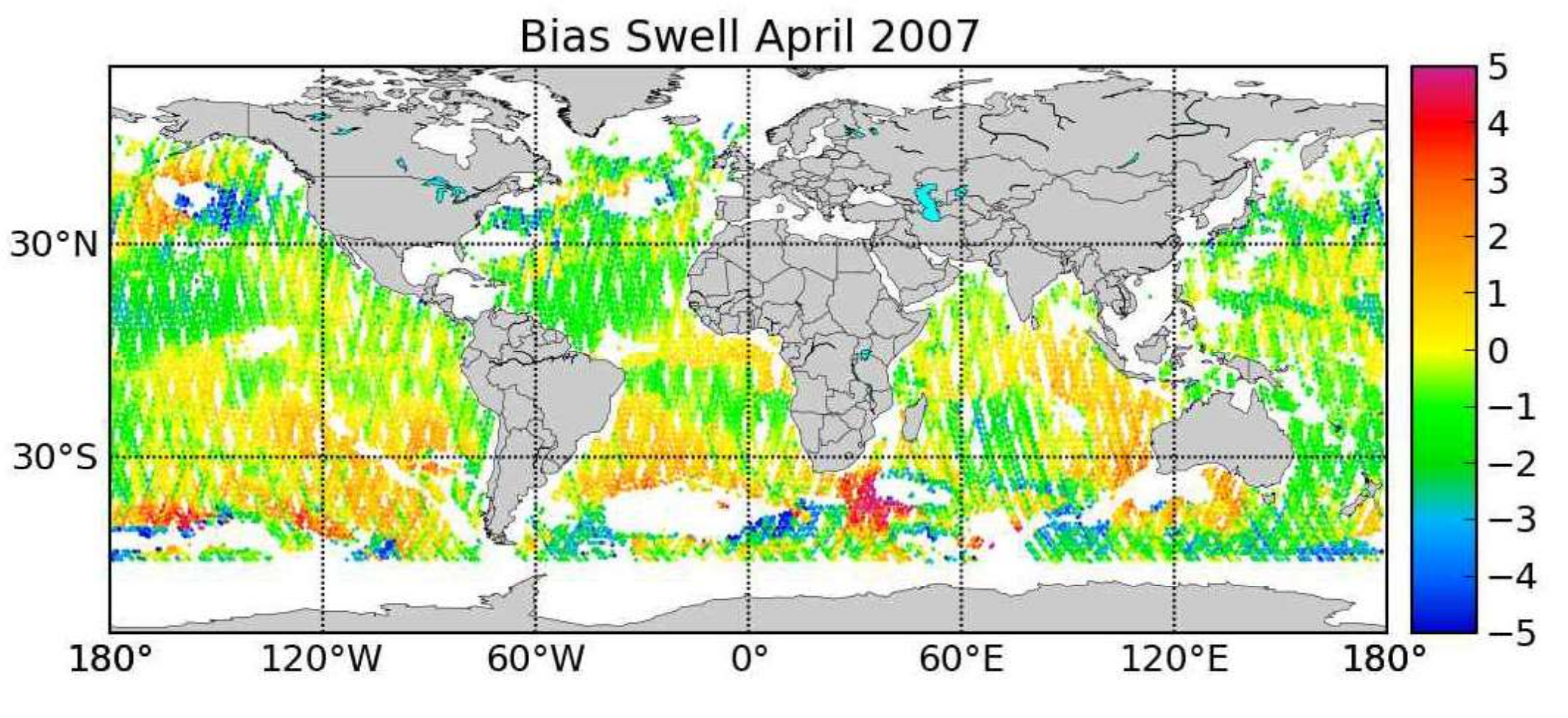} \\[1.cm]
    \includegraphics[width=6.9cm]{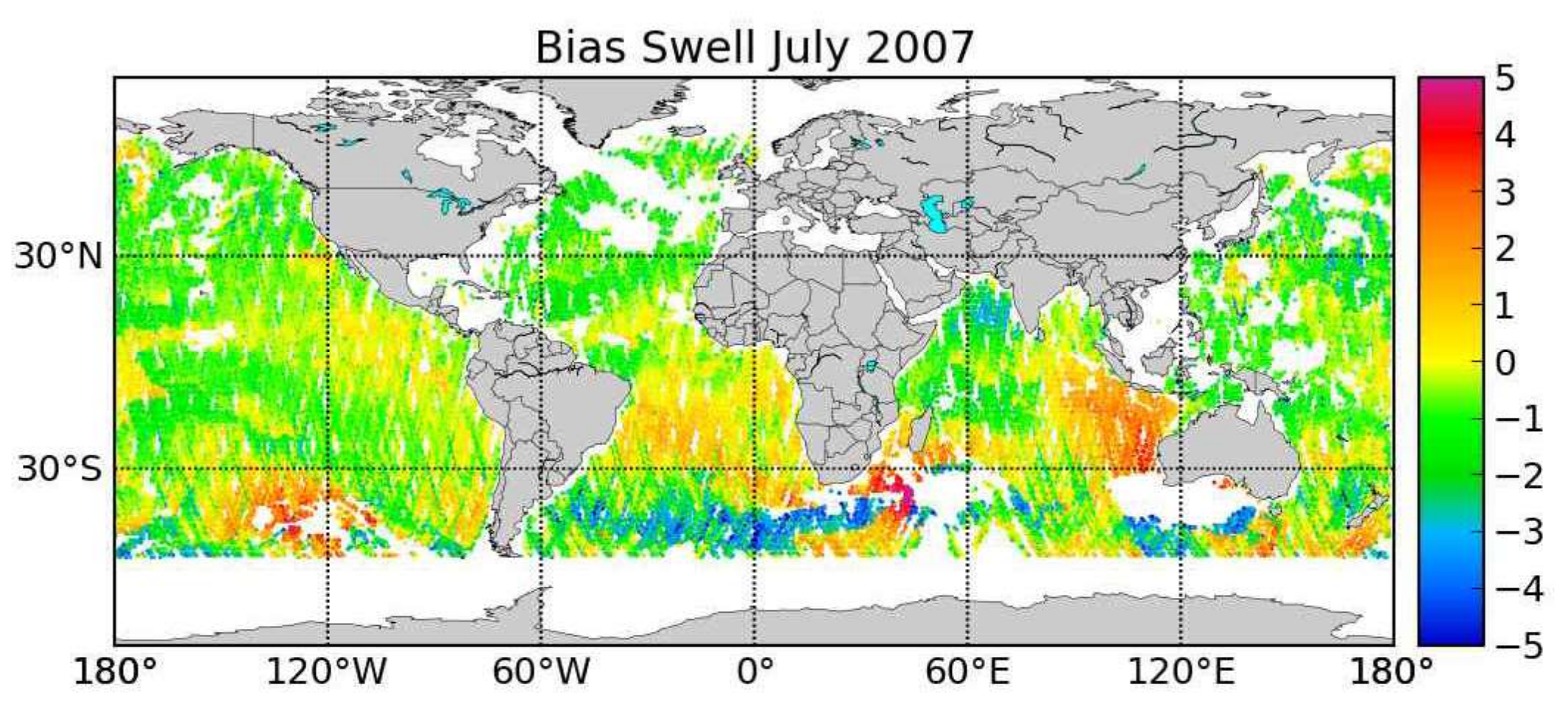} &
    \includegraphics[width=7.cm]{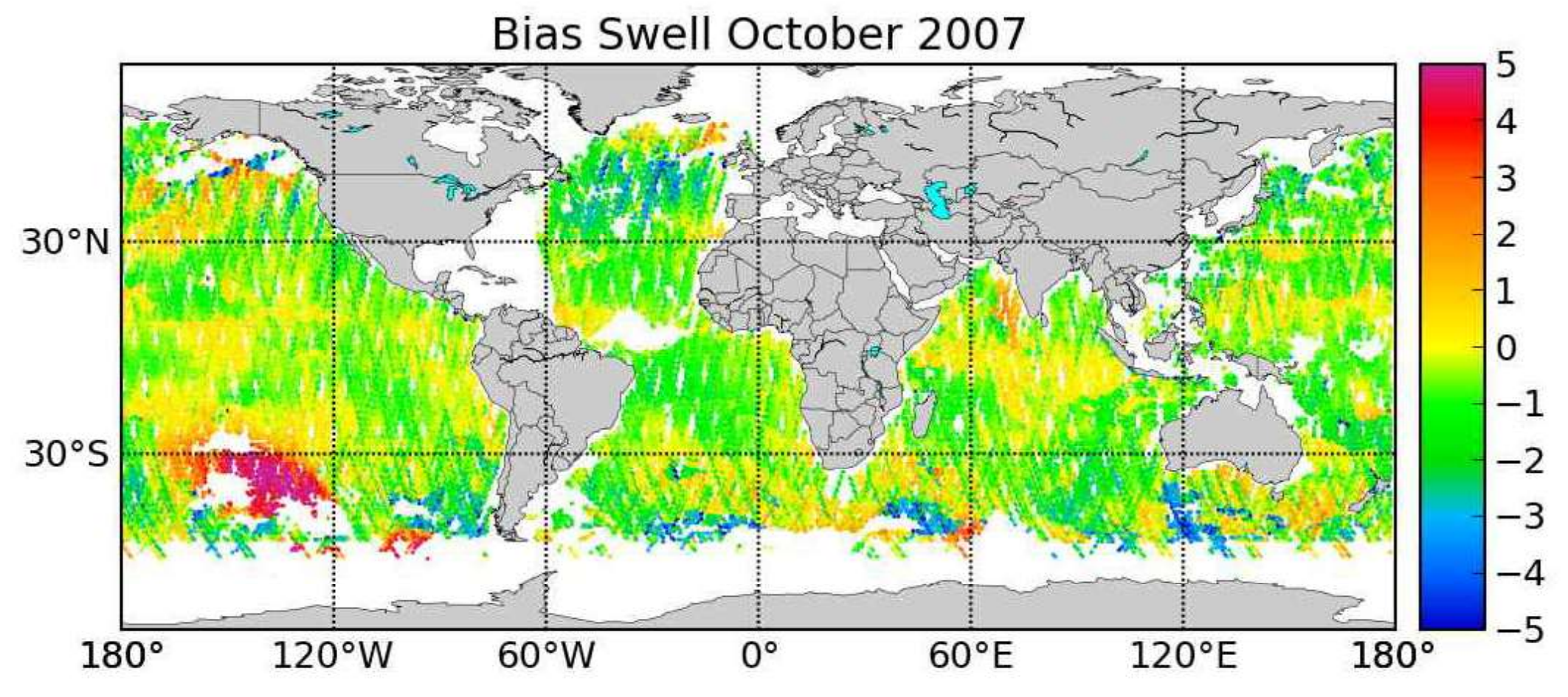}
    \end{array}$
    \caption{Global bias for all four seasons of 2007.}
    \label{oldfig2}
    \end{center}
\end{figure*}

\begin{figure*}[h]
    \begin{center}$
   \begin{array}{cc}
    \includegraphics[width=6.9cm]{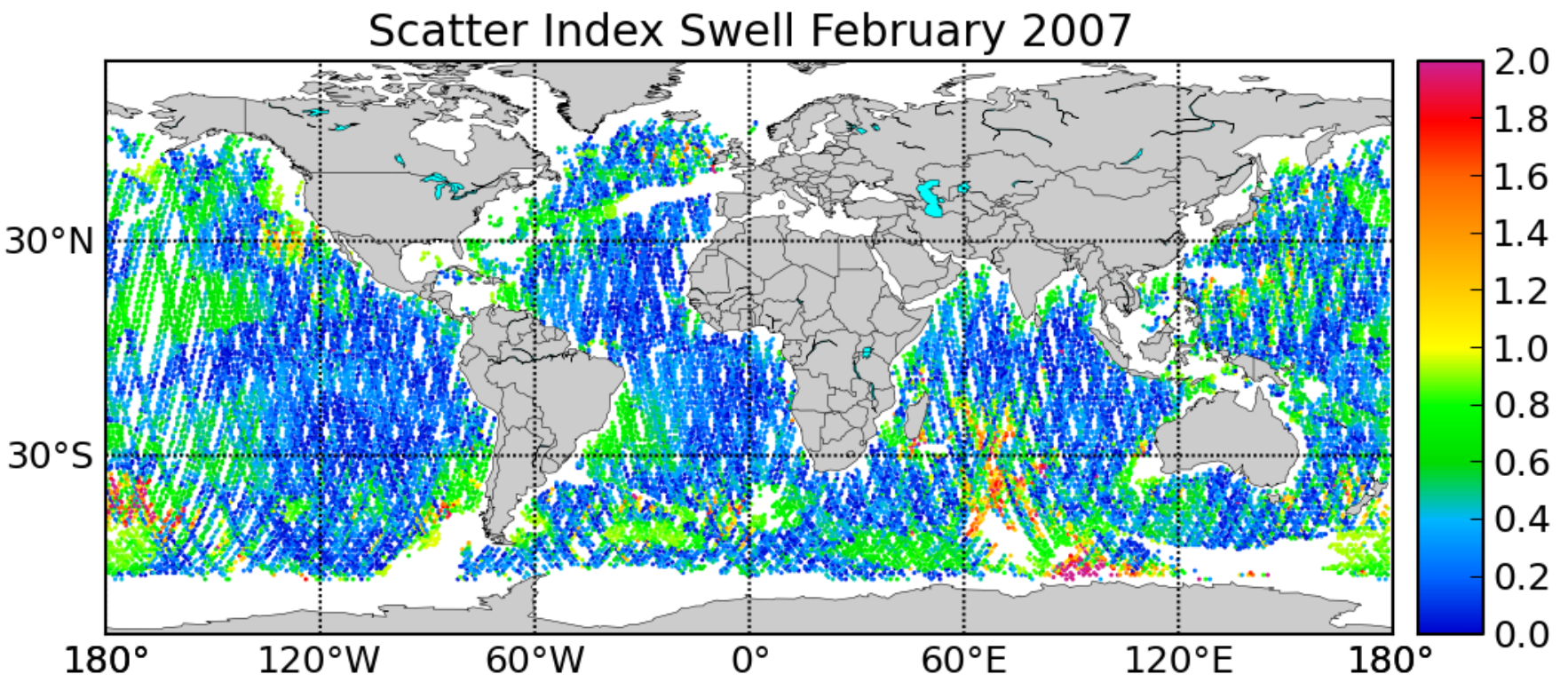} &
    \includegraphics[width=7.cm]{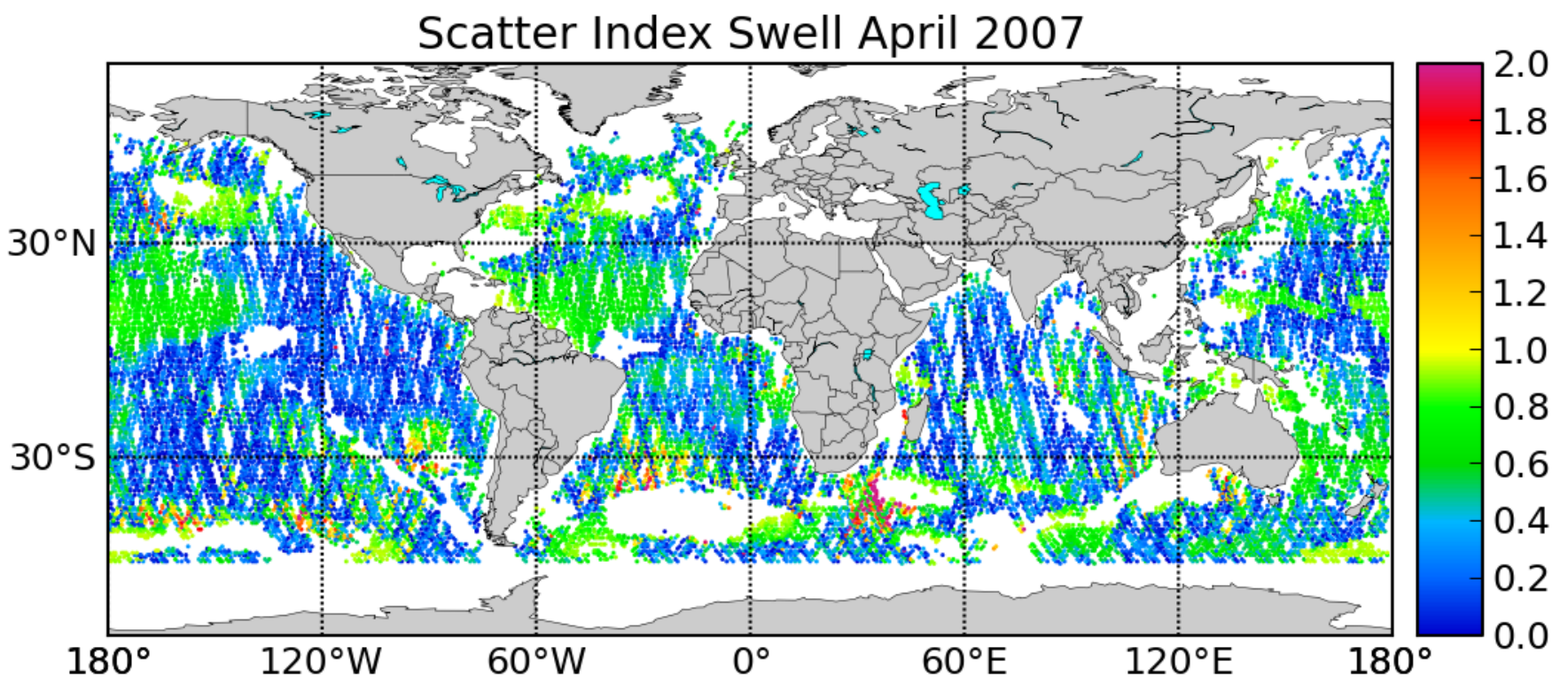} \\[1.cm]
    \includegraphics[width=6.9cm]{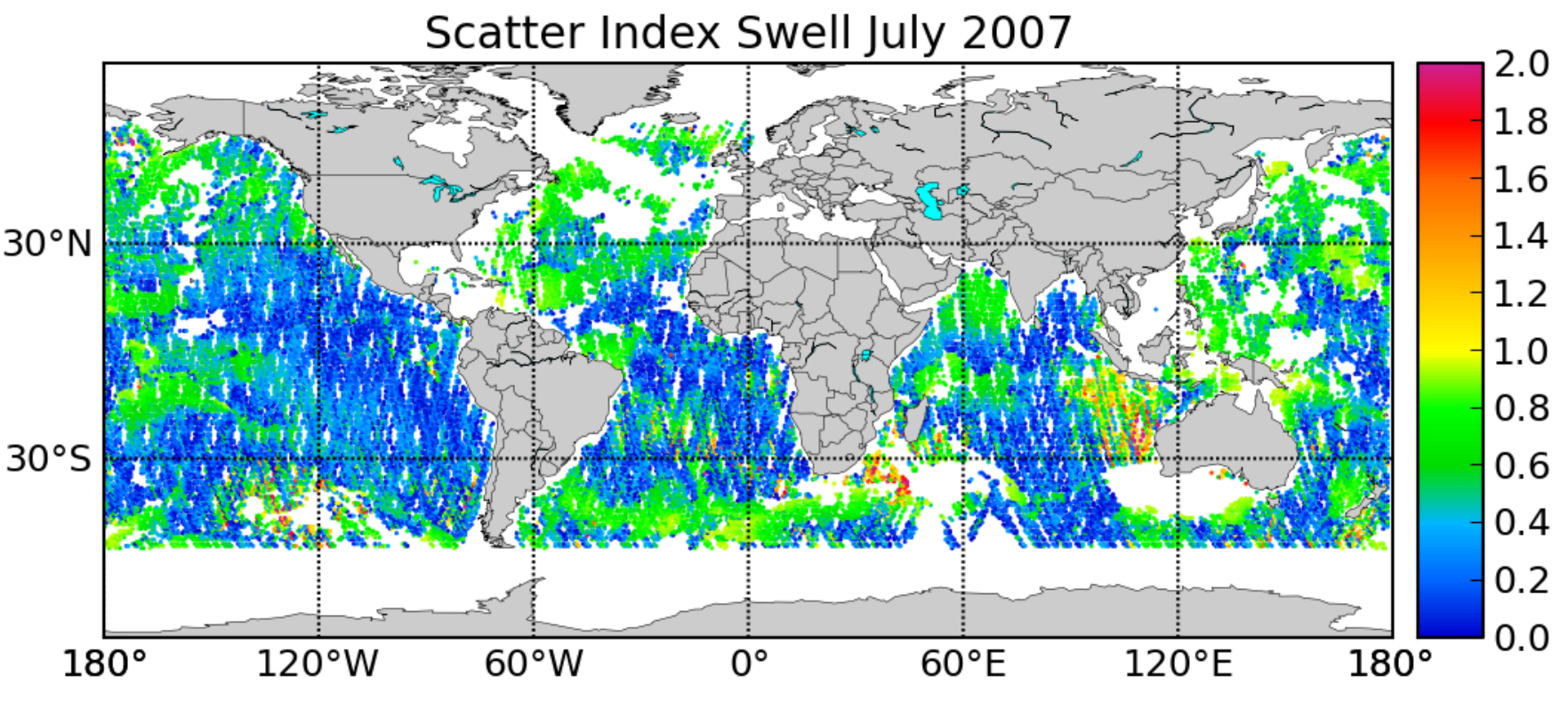} &
    \includegraphics[width=7.cm]{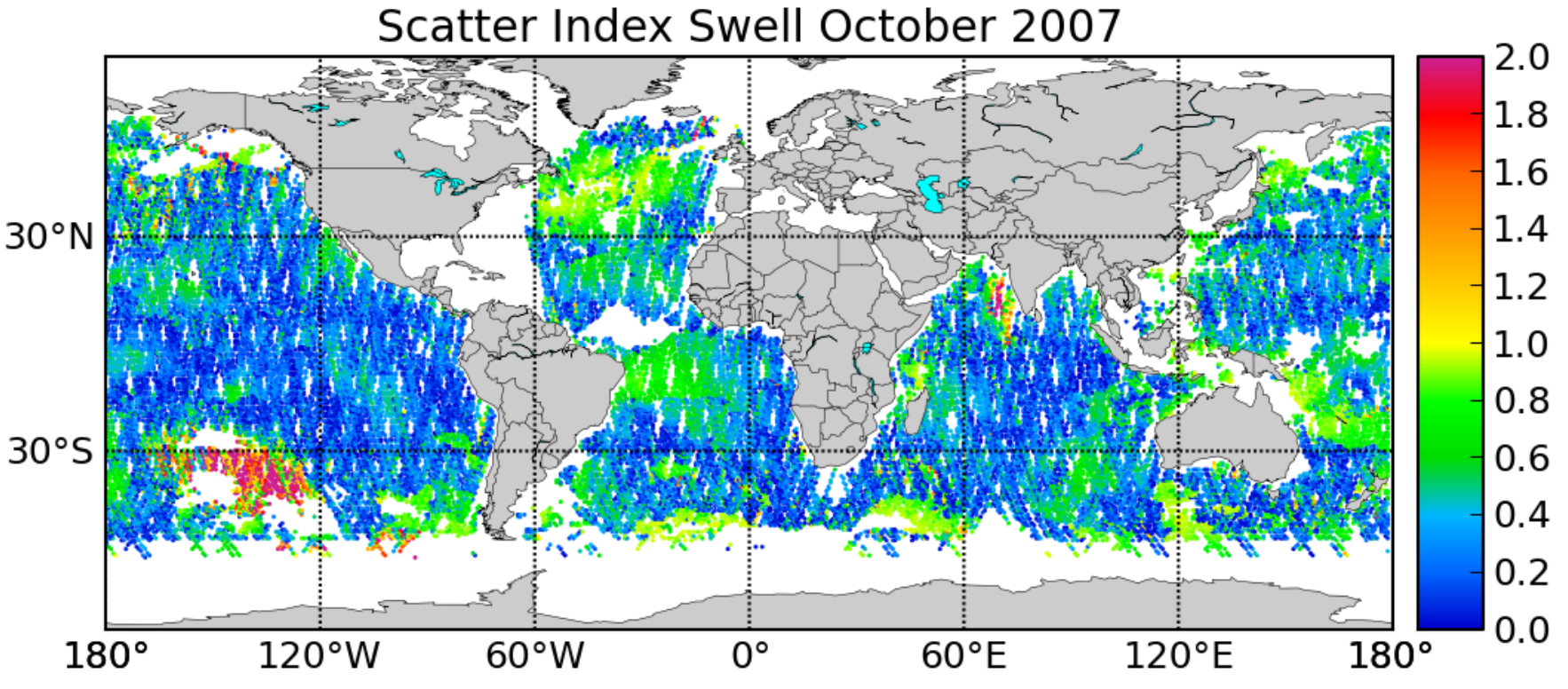}
    \end{array}$
    \caption{Global scattering index for all four seasons of 2007.}
    \label{oldfig3}
    \end{center}
\end{figure*}

Satellite measurements show very clearly high swell heights in the Northwestern Indian Ocean in the North Hemisphere late spring (June) and summer (figure~\ref{fig:julout-swell}). This can be explained by strong southwest winds that develop near the coast of Africa and is associated with the Asian summer monsoon~\cite{Young1999}. The negative bias of July in this region, shown in figure~\ref{oldfig2}, indicates the model did not accurately pick up this phenomenon.
The largest bias in the comparison satellite-model, presented in figure~\ref{oldfig2}, were observed at high latitudes, specially in the SH. 
In figure~\ref{oldfig3}, we see that the scatter index is below 0.5, almost everywhere. In some regions, these values were close to 1. Higher values (greater than 1) were observed in the North Atlantic, South Pacific and Northern Australia.

The bias we found presents complex characteristics for the year analysed. We could infer that for the months between February and April, the highest bias, meaning overestimation by the model, are predominantly positive with a remarkable exception in regions of the Atlantic in latitude range 40-45 S and also in 40-45 N, including the
North Pacific. In the months of July and October, the same extra-tropical range show underestimation by the model in the Atlantic, and South Indian Ocean. In the tropics, our results show less intense but more extensive areas with negative bias, and thus model underestimation of swell heights apparently contradicting previous findings of systematic overestimation of wave heights by numerical wave models~\cite{Rascle2008}.
This may be related to the fact that in the algorithm used in this work, when more than one satellite measurement is found in the same cell of the model mesh for the same period of time, taken here as one month, the highest value is selected and a new value computed by 
the model is also picked up within the shortest time tolerance. Thus, this procedure is 
done without compromising the nearness in time for each two values compared. 
\clearpage
%
% *********************************************************
\section{Conclusions} \label{sec:end}
% *********************************************************
In this work an algorithm for representing irregular satellite wave data
on an uniform mesh has been introduced. The novelty of this procedure is that both the satellite and model data values correspond to time instances, not means or averages, at the same point in space. The graphical representation of these data at points rather than grid cells corroborates with this approach and complements it.

ENVISAT swell heights, obtained through the ASAR sensor have been compared  with swell computed by the WaveWatchIII. A short-scale simulation of the wave model was performed for the entire year of 2007 and results for a month of each season have been analysed through bias and scatter index.

Known features of the global wave climate have been corroborated as well as specific findings for the year studied. In special, higher latitude regions where the model underestimate the measured swell heights have been identified. Further, it has been  noticed that in the tropics, less intense but more extensive areas also present negative bias, and thus model underestimation of swell heights. These findings should provide additional insights for the ocean wave modelling community and indication of future model improvements and calibration.

With this new picture it became clear that there are spatial discrepancies. The energy in the tropical oceans is concentrated in the low frequency, longer swell waves due to the weaker wind pattern in this part of the world. In contrast, higher latitudes are windier and most of the wave energy is in its higher frequencies. Therefore, the content of energy in the tropics is biased towards swell waves, which are more effectively measured by ASAR. Here we show that in the tropics the discrepancies are more evident, which raises questions about the performance of the model (or ASAR retrievals) during the period used for our validation.

The features of the satellite representation introduced here must be 
further investigated and extended to other physical variables such as
wave period and direction and a similar study as the one presented here should be carried out using a more comprehensive dataset, incorporating several years. However, the results show the potential of the proposed technique where several discrepancies between  satellite data and the output from a well-known numerical model were pinpointed. 

\section*{Acknowledgements}
HP acknowledges financial support by CAPES, Brazil, through the Project Edital Capes Ci\^encias do Mar 09/2009.
This study was financed in part by the Coordenação de Aperfeiçoamento
de Pessoal de Nível Superior - Brasil (CAPES) - Finance Code 001. Leandro Farina acknowledges support from the project ROAD-BESM (REGIONAL OCEANIC AND ATMOSPHERIC DOWNSCALING CAPES (88881.146048/2017-01)).

\section*{Annex}
The algorithm implemented in python and used in this work is reproduced below.
\begin{verbatim}
#Hs=satelite wave height
#swell_total=model swell height

#Procedure 1

At1=np.zeros(np.shape(lons1))

""" lon e lat devem possuir um valor a mais do que i e j """
                    
for i in range(0,233):    
    for j in range(0,600):
        for k in range(len(Hs)):
            if PtoX[k]>lons[j] and PtoX[k]<lons[j+1]:
                if PtoY[k]>lats[i] and PtoY[k]<lats[i+1]:
                    if At1[i,j]<Hs[k]:
                        At1[i,j]=Hs[k]
                    
                    
l=np.nonzero(At1==0)
At1[l]=nan

del l

Hsat=np.copy(At1)

#Procedure 2

Hmod=np.copy(Nan)

for ti in range(0,233):
  for tj in range(0,600):
    for tm in range(len(datas_python_jan)-1):
      for tm2 in range(len(datas_python_jan)):
         if datas_sat[ti,tj]>datas_python_jan[tm] and datas_sat[ti,tj]<datas_python_jan[tm+1]:
               Hmod[ti,tj]=swell_total[tm2,ti,tj]
\end{verbatim}

\clearpage

\bibliographystyle{plain}
\bibliography{paper2}

\end{document}